\RequirePackage{lineno}
\documentclass[aps,prl,twocolumn,showpacs,superscriptaddress,groupedaddress]{revtex4}
\usepackage{graphicx}
\usepackage{subfigure}
\usepackage{bm}
\usepackage{exscale}
\usepackage{amssymb}
\usepackage{ifthen}
\usepackage{latexsym}
\usepackage{makeidx}
\usepackage{fancyhdr}
\usepackage{syntonly}
\usepackage{CJK}
\usepackage{amsmath}
\usepackage{dcolumn}
\usepackage{tikz}
\usepackage{float}
\usepackage{enumitem}
\usepackage{footmisc}

\usetikzlibrary{trees}
\usetikzlibrary{decorations.pathmorphing}
\usetikzlibrary{decorations.markings}

\newcommand{\pmcs}{\textsc{PMCS}\,}
\newcommand{\resbos}{{\sc resbos}~}
\newcommand{\pythia}{{\sc pythia}~}

\newcommand{\pTW}{\ensuremath{p_T^W}}

\newcommand{\pTV}{\ensuremath{p_T^V}}
\newcommand{\dzero}{\ensuremath{\mbox{D0}}~}
\newcommand{\pte}{\ensuremath{p_T^e}}

\newcommand{\met}{\ensuremath{{\slash\kern-.7emE}_{T}}}
\newcommand{\mt}{\ensuremath{m_T}}
\newcommand{\vut}{\ensuremath{\vec{u}_T}}
\newcommand{\vmet}{\ensuremath{\vec{\slash\kern-.7emE}_{T}}}

\begin{document}
	
	
	
		\noindent%
		\mbox{FERMILAB-PUB-20-354-E \hspace{15mm}{\em Published in Phys. Rev. D as
				DOI: 10.1103/PhysRevD.103.012003}}

	
	
	\title{\bf Study of the normalized transverse momentum distribution of $W$ bosons produced in $p \bar p$ collisions at $\sqrt{s}=1.96$ TeV}
	\affiliation{LAFEX, Centro Brasileiro de Pesquisas F\'{i}sicas, Rio de Janeiro, RJ 22290, Brazil}
	\affiliation{Universidade do Estado do Rio de Janeiro, Rio de Janeiro, RJ 20550, Brazil}
	\affiliation{Universidade Federal do ABC, Santo Andr\'e, SP 09210, Brazil}
	\affiliation{University of Science and Technology of China, Hefei 230026, People's Republic of China}
	\affiliation{Universidad de los Andes, Bogot\'a, 111711, Colombia}
	\affiliation{Charles University, Faculty of Mathematics and Physics, Center for Particle Physics, 116 36 Prague 1, Czech Republic}
	\affiliation{Czech Technical University in Prague, 116 36 Prague 6, Czech Republic}
	\affiliation{Institute of Physics, Academy of Sciences of the Czech Republic, 182 21 Prague, Czech Republic}
	\affiliation{Universidad San Francisco de Quito, Quito 170157, Ecuador}
	\affiliation{LPC, Universit\'e Blaise Pascal, CNRS/IN2P3, Clermont, F-63178 Aubi\`ere Cedex, France}
	\affiliation{LPSC, Universit\'e Joseph Fourier Grenoble 1, CNRS/IN2P3, Institut National Polytechnique de Grenoble, F-38026 Grenoble Cedex, France}
	\affiliation{CPPM, Aix-Marseille Universit\'e, CNRS/IN2P3, F-13288 Marseille Cedex 09, France}
	\affiliation{LAL, Univ. Paris-Sud, CNRS/IN2P3, Universit\'e Paris-Saclay, F-91898 Orsay Cedex, France}
	\affiliation{LPNHE, Universit\'es Paris VI and VII, CNRS/IN2P3, F-75005 Paris, France}
	\affiliation{IRFU, CEA, Universit\'e Paris-Saclay, F-91191 Gif-Sur-Yvette, France}
	\affiliation{IPHC, Universit\'e de Strasbourg, CNRS/IN2P3, F-67037 Strasbourg, France}
	\affiliation{IPNL, Universit\'e Lyon 1, CNRS/IN2P3, F-69622 Villeurbanne Cedex, France and Universit\'e de Lyon, F-69361 Lyon CEDEX 07, France}
	\affiliation{III. Physikalisches Institut A, RWTH Aachen University, 52056 Aachen, Germany}
	\affiliation{Physikalisches Institut, Universit\"at Freiburg, 79085 Freiburg, Germany}
	\affiliation{II. Physikalisches Institut, Georg-August-Universit\"at G\"ottingen, 37073 G\"ottingen, Germany}
	\affiliation{Institut f\"ur Physik, Universit\"at Mainz, 55099 Mainz, Germany}
	\affiliation{Ludwig-Maximilians-Universit\"at M\"unchen, 80539 M\"unchen, Germany}
	\affiliation{Panjab University, Chandigarh 160014, India}
	\affiliation{Delhi University, Delhi-110 007, India}
	\affiliation{Tata Institute of Fundamental Research, Mumbai-400 005, India}
	\affiliation{University College Dublin, Dublin 4, Ireland}
	\affiliation{Korea Detector Laboratory, Korea University, Seoul, 02841, Korea}
	\affiliation{CINVESTAV, Mexico City 07360, Mexico}
	\affiliation{Nikhef, Science Park, 1098 XG Amsterdam, the Netherlands}
	\affiliation{Radboud University Nijmegen, 6525 AJ Nijmegen, the Netherlands}
	\affiliation{Joint Institute for Nuclear Research, Dubna 141980, Russia}
	\affiliation{Institute for Theoretical and Experimental Physics, Moscow 117259, Russia}
	\affiliation{Moscow State University, Moscow 119991, Russia}
	\affiliation{Institute for High Energy Physics, Protvino, Moscow region 142281, Russia}
	\affiliation{Petersburg Nuclear Physics Institute, St. Petersburg 188300, Russia}
	\affiliation{Instituci\'{o} Catalana de Recerca i Estudis Avan\c{c}ats (ICREA) and Institut de F\'{i}sica d'Altes Energies (IFAE), 08193 Bellaterra (Barcelona), Spain}
	\affiliation{Uppsala University, 751 05 Uppsala, Sweden}
	\affiliation{Taras Shevchenko National University of Kyiv, Kiev, 01601, Ukraine}
	\affiliation{Lancaster University, Lancaster LA1 4YB, United Kingdom}
	\affiliation{Imperial College London, London SW7 2AZ, United Kingdom}
	\affiliation{The University of Manchester, Manchester M13 9PL, United Kingdom}
	\affiliation{University of Arizona, Tucson, Arizona 85721, USA}
	\affiliation{University of California Riverside, Riverside, California 92521, USA}
	\affiliation{Florida State University, Tallahassee, Florida 32306, USA}
	\affiliation{Fermi National Accelerator Laboratory, Batavia, Illinois 60510, USA}
	\affiliation{University of Illinois at Chicago, Chicago, Illinois 60607, USA}
	\affiliation{Northern Illinois University, DeKalb, Illinois 60115, USA}
	\affiliation{Northwestern University, Evanston, Illinois 60208, USA}
	\affiliation{Indiana University, Bloomington, Indiana 47405, USA}
	\affiliation{Purdue University Calumet, Hammond, Indiana 46323, USA}
	\affiliation{University of Notre Dame, Notre Dame, Indiana 46556, USA}
	\affiliation{Iowa State University, Ames, Iowa 50011, USA}
	\affiliation{University of Kansas, Lawrence, Kansas 66045, USA}
	\affiliation{Louisiana Tech University, Ruston, Louisiana 71272, USA}
	\affiliation{Northeastern University, Boston, Massachusetts 02115, USA}
	\affiliation{University of Michigan, Ann Arbor, Michigan 48109, USA}
	\affiliation{Michigan State University, East Lansing, Michigan 48824, USA}
	\affiliation{University of Mississippi, University, Mississippi 38677, USA}
	\affiliation{University of Nebraska, Lincoln, Nebraska 68588, USA}
	\affiliation{Rutgers University, Piscataway, New Jersey 08855, USA}
	\affiliation{Princeton University, Princeton, New Jersey 08544, USA}
	\affiliation{State University of New York, Buffalo, New York 14260, USA}
	\affiliation{University of Rochester, Rochester, New York 14627, USA}
	\affiliation{State University of New York, Stony Brook, New York 11794, USA}
	\affiliation{Brookhaven National Laboratory, Upton, New York 11973, USA}
	\affiliation{Langston University, Langston, Oklahoma 73050, USA}
	\affiliation{University of Oklahoma, Norman, Oklahoma 73019, USA}
	\affiliation{Oklahoma State University, Stillwater, Oklahoma 74078, USA}
	\affiliation{Oregon State University, Corvallis, Oregon 97331, USA}
	\affiliation{Brown University, Providence, Rhode Island 02912, USA}
	\affiliation{University of Texas, Arlington, Texas 76019, USA}
	\affiliation{Southern Methodist University, Dallas, Texas 75275, USA}
	\affiliation{Rice University, Houston, Texas 77005, USA}
	\affiliation{University of Virginia, Charlottesville, Virginia 22904, USA}
	\affiliation{University of Washington, Seattle, Washington 98195, USA}
	\author{V.M.~Abazov} \affiliation{Joint Institute for Nuclear Research, Dubna 141980, Russia}
	\author{B.~Abbott} \affiliation{University of Oklahoma, Norman, Oklahoma 73019, USA}
	\author{B.S.~Acharya} \affiliation{Tata Institute of Fundamental Research, Mumbai-400 005, India}
	\author{M.~Adams} \affiliation{University of Illinois at Chicago, Chicago, Illinois 60607, USA}
	\author{T.~Adams} \affiliation{Florida State University, Tallahassee, Florida 32306, USA}
	\author{J.P.~Agnew} \affiliation{The University of Manchester, Manchester M13 9PL, United Kingdom}
	\author{G.D.~Alexeev} \affiliation{Joint Institute for Nuclear Research, Dubna 141980, Russia}
	\author{G.~Alkhazov} \affiliation{Petersburg Nuclear Physics Institute, St. Petersburg 188300, Russia}
	\author{A.~Alton$^{a}$} \affiliation{University of Michigan, Ann Arbor, Michigan 48109, USA}
	\author{A.~Askew} \affiliation{Florida State University, Tallahassee, Florida 32306, USA}
% ...
	%
	%
	\collaboration{The D0 Collaboration\footnote{with visitors from
			$^{a}$Augustana College, Sioux Falls, SD 57197, USA,
			$^{b}$The University of Liverpool, Liverpool L69 3BX, UK,
			$^{c}$Deutshes Elektronen-Synchrotron (DESY), Notkestrasse 85, Germany,
			$^{d}$CONACyT, M-03940 Mexico City, Mexico,
			$^{e}$SLAC, Menlo Park, CA 94025, USA,
			$^{f}$University College London, London WC1E 6BT, UK,
			$^{g}$Centro de Investigacion en Computacion - IPN, CP 07738 Mexico City, Mexico,
			$^{h}$Universidade Estadual Paulista, S\~ao Paulo, SP 01140, Brazil,
			$^{i}$Karlsruher Institut f\"ur Technologie (KIT) - Steinbuch Centre for Computing (SCC),
			D-76128 Karlsruhe, Germany,
			$^{j}$Office of Science, U.S. Department of Energy, Washington, D.C. 20585, USA,
			$^{k}$Kiev Institute for Nuclear Research (KINR), Kyiv 03680, Ukraine,
			$^{l}$University of Maryland, College Park, MD 20742, USA,
			$^{m}$European Orgnaization for Nuclear Research (CERN), CH-1211 Geneva, Switzerland,
			$^{n}$Purdue University, West Lafayette, IN 47907, USA,
			$^{o}$Institute of Physics, Belgrade, Belgrade, Serbia,
			and
			$^{p}$P.N. Lebedev Physical Institute of the Russian Academy of Sciences, 119991, Moscow, Russia.
			$^{\ddag}$Deceased.
	}} \noaffiliation
	\vskip 0.25cm
	      
	\date{\today}
	
	\begin{abstract}
		{\normalsize We present a study of the normalized transverse momentum distribution of $W$ bosons produced in $p \bar p$ collisions, using data corresponding to an integrated luminosity of 4.35~fb$^{-1}$ collected with the \dzero detector at the Fermilab Tevatron collider at $\sqrt{s}=1.96$~TeV. The measurement focuses on the transverse momentum region below 15 GeV, which is of special interest for electroweak precision measurements; it relies on the same detector calibration methods which were used for the precision measurement of the $W$ boson mass. The measured distribution is compared to different QCD predictions and a procedure is given to allow the comparison of any further theoretical models to the D0 data.}
	\end{abstract}
	
	\maketitle

	
	
	~\\
	\begin{center}
		\bf I. INTRODUCTION
	\end{center}
	~\\
	
	The production of $V=(W/Z)$ bosons in hadron collisions is described by perturbative quantum chromodynamics (QCD). At leading order, QCD predicts no transverse momentum of the $W$ or $Z$ boson ($\pTV$) with respect to the beam direction \cite{ATLASReview}. However, this changes when including higher order corrections, so that significant $\pTV$ can arise from the emission of partons in the initial state as well as from the intrinsic transverse momentum of the initial-state partons in the proton. The $\pTV$ spectrum at low transverse momentum can be described using soft-gluon resummation \cite{Collins:1984kg,Balazs:1997xd, Becher:2010tm, Catani:2015vma, Ebert:2016gcn, Bizon:2017rah}, parton shower approaches \cite{Bellm:2015jjp,Sjostrand:2014zea,Bothmann:2019yzt}, and non-perturbative calculations \cite{NPQCD, NPFunc} to account for the intrinsic transverse momentum of partons. In the non-perturbative approach~\cite{NPQCD, NPFunc}, a function is introduced as a form factor in order to make the QCD calculation convergent when $\pTV\rightarrow0^{+}$. The values of the parameters in the non-perturbative function can only be extracted from the measurement of the $\pTV$ distribution. Knowledge of the $\pTV$ spectrum is not only important for testing perturbative QCD predictions and constraining models of non-perturbative approaches, but also for the measurement of electroweak parameters such as the $W$ boson mass. In the latter case, it is especially important to model the $\pTW$ spectrum correctly in the low $p_T$ region. 
	
	The transverse momentum spectrum of the $Z$ boson has been measured to high precision at various energies, both at the Tevatron \cite{DZeroZPt1, DZeroZPt2, DZeroZPt3, CDFZPt}  and the LHC \cite{ATLASZPt1, ATLASZPt2, ATLASZPt3, CMSZPt1, CMSZPt2, CMSPt}. This precision is enabled by the fact that leptonically-decaying $Z$ bosons can be easily reconstructed from the two charged leptons in the final state. The situation is different for the $W$ boson as the neutrino escapes detection and hadronic decays have large backgrounds. The $\pTW$ must therefore be estimated from the reconstructed hadronic recoil of the event. The hadronic recoil is only an approximation of $p_T(W)$ as it is significantly affected by the number of simultaneous hadron collisions in the recorded event and by the non-linear energy response of the detector for low energy hadrons. 
	
	The $\pTW$ distribution was previously measured at the Tevatron at $\sqrt{s}=1.8$ TeV~\cite{DZeroWPt1, DZeroWPt2}, and at the LHC at $\sqrt{s}=7$ and $8$~TeV~\cite{ATLASWPt, CMSPt}. This study is the first $\pTW$ analysis at $\sqrt{s}=1.96$~TeV. In this paper, we analyze data corresponding to an integrated luminosity of 4.35 fb$^{-1}$ collected by the \dzero detector at the Fermilab Tevatron collider. These data were also used for the $W$ boson mass measurement in Ref.~\cite{WMass}. This study concentrates on the low $\pTW$ region and resolves the peak near $\pTW=4$~GeV, unlike the LHC measurements of Refs.~\cite{ATLASWPt, CMSPt} where the sizes of the first bin are 8 GeV and 7.5 GeV, respectively. In addition, we study the transverse momentum of $W$ bosons in the case where the production is dominated by valence quarks, unlike the situation at the LHC which involves sea quarks. Typical Bjorken $x$-values for $W$ boson production at the Tevatron (LHC) are 0.05 (0.015) \cite{ATLASReview}. 
	
	This paper is structured as follows: after a short description of the \dzero detector, the event selection, the calibration procedure, and the basic comparison plots between data and simulation are presented. This is followed by a description of the analysis procedure. After a discussion of the systematic uncertainties, the final results are presented and compared with several models of $W$ boson production and parton distribution functions. Finally, a fast folding procedure is introduced in the appendix, which can be used to compare our result to other theoretical predictions while properly accounting for the detector response.
	
	~\\
	\begin{center}
		\bf II. THE D0 DETECTOR
	\end{center}
	~\\
	
    The D0 detector~\cite{d0-detector} comprises a central tracking system, a calorimeter, and a muon system.  The analysis uses a cylindrical coordinate system with the $z$ axis along the beam axis in the proton direction. Angles $\theta$ and $\phi$ are the polar and azimuthal angles, respectively. Pseudorapidity is defined as $\eta=-\ln [\tan(\theta/2)]$ where $\theta$ is measured with respect to the interaction vertex. We define $\eta_{\text{det}}$ as the pseudorapidity measured with respect to the center of the detector.  The central tracking system consists of a silicon microstrip tracker (SMT) and a scintillating fiber tracker, both located within a 1.9~T superconducting solenoid magnet and optimized for tracking and vertexing for $|\eta_{\text{det}}|<2.5$. Outside the solenoid, liquid argon and uranium calorimeters provide energy measurement, with a central calorimeter (CC) that covers $|\eta_\text{det}|\le 1.05$, and two end calorimeters (EC) that extend coverage to $|\eta_\text{det}|<4.2$.  The muon system located outside the calorimeter consists of drift tubes and scintillators before and after 1.8~T iron toroid magnets and provides coverage for $|\eta_\text{det}|<2.0$. Muons are identified and their momenta are measured using information from both the tracking system and the muon system. The solenoid and toroid polarities are reversed every two weeks on average during the periods of data-taking.
	
	~\\
	\begin{center}
		\bf III. EVENT SAMPLES AND EVENT SELECTION
	\end{center}
	~\\

	The present analysis builds on the techniques developed in Refs.~\cite{WMass} and \cite{D0NewW} for the measurement of the $W$ boson mass. Events are selected using a trigger requiring at least one electromagnetic (EM) cluster found in the CC, with the transverse energy threshold varying from 25 to 27~GeV depending on run conditions.  The offline selection of candidate $W$~boson events is the same as used in Ref.~\cite{WMass}. We require candidate electrons to be matched in $(\eta,\phi)$ space to a track including at least one SMT hit. The electron three-momentum vector magnitude is defined by the cluster energy, and the direction is defined by the track.
	
	We require the presence of an electron with $\pte>25$~GeV and $|\eta^e|<1.05$ that passes shower shape and isolation requirements. Here $p_T^e$ is the magnitude of the transverse momentum of the electron, $\vec{p}_{T}^{~e}$, and $\eta^e$ is the pseudorapidity of the electron. The event must satisfy $~\met > 25$~GeV, $u_T < 15$~GeV, and $50 < \mt < 200$~GeV. Here, the hadronic recoil $\vec{u}_T$ is the vector sum of the transverse component of the energies measured in calorimeter cells excluding those associated with the reconstructed electron, and $u_T$ is its magnitude. The relation $~\vmet=-(\vec{p}_T^{~e}+\vut)$ defines the missing transverse energy approximating the transverse momentum of the neutrino, and $\mt$ is the transverse mass defined as $\mt = \sqrt{2p_{T}^{e}\met(1-\cos \Delta \phi)}$, where $\Delta \phi$ is the azimuthal opening angle between $\vec{p}_{T}^{~e}$ and $\vmet$. 
	This selection yields 1\,677\,394 candidate $W\to e\nu$ events. 
	
	The $Z\to ee$ events were used extensively to calibrate the detector response~\cite{D0NewW, WMass}, and they are also used in this study. These events are required to have two EM clusters satisfying the $W$ candidate cluster requirements above, except that one of the two clusters may be reconstructed within an EC $(1.5<|\eta|<2.5)$. The associated tracks must be of opposite curvature.  The $Z$ boson events must also have $u_T < 15$~GeV and $70 \le m_{ee} \le 110$~GeV, where $m_{ee}$ is the invariant mass of the electron-positron pair. 
	
	The \resbos~\cite{Balazs:1997xd} event generator, combined with {\sc photos}~\cite{photos}, is used as a baseline simulation for the kinematics of $W$ and $Z$ boson production and decay. \resbos is a next-to-leading order event generator including next-to-next-to-leading logarithm resummation of soft gluons~\cite{Collins:1984kg}, and {\sc photos} generates up to two final state radiated photons. At low transverse momentum ($\pTV < 10$ GeV), multiple soft gluon emissions dominate the cross section and a soft-gluon resummation formalism is used to make QCD predictions. This technique was first developed by Collins, Soper, and Sterman (CSS) \cite{Collins:1984kg} and is currently implemented using a parametric function introduced by Brock, Landry, Nadolsky and Yuan (BLNY)~\cite{g2} based on three non-perturbative parameters $g_1$, $g_2$ and $g_3$. In the
	kinematic region of this measurement, the $\pTW$ distribution is insensitive to $g_3$, but can be used to constrain $g_1$ and $g_2$. The baseline simulation relies on the CTEQ6.6~\cite{cteq66} PDF set, as well as setting the non-perturbative parameters to the following values from Ref.~\cite{g2}: $g_1 = ~0.21$~GeV$^2$, $g_2 = ~0.68$~GeV$^2$, and $g_3 = -0.60$~GeV$^2$.
    
	
	We compare our measurement with predictions from various Monte Carlo (MC) simulations (\resbos and \pythia\cite{Sjostrand:2014zea}), different PDF sets (CT14HERA2NNLO~\cite{CT14HERA2,CT14HERA2NNLO}, CTEQ6L1~\cite{CTEQ6L1}, MSTW2008LO~\cite{MSTW2008} and MRST LO~\cite{MRST}) and two non-perturbative functional forms (BLNY and the transverse momentum dependent TMD-BLNY~\cite{TMDBLNY}):
	\newline
	
	\begin{enumerate}[nosep, noitemsep, leftmargin=4mm]
		\setlength{\itemsep}{0pt}
		\setlength{\parsep}{0pt}
		\setlength{\parskip}{0pt}
		\setlength{\labelwidth}{0pt}
		\setlength{\labelindent}{0pt}
		{\small
			\item \resbos~(Version~CP020811)+BLNY+CTEQ6.6
			\item \resbos~(Version~CP112216)+TMD-BLNY\\+CT14HERA2NNLO
			\item \pythia8+CT14HERA2NNLO
			\item \pythia8+ATLAS~MB~A2Tune~\cite{ATL-PHYS-PUB-2012-003}+CTEQ6L1
			\item \pythia8+ATLAS~MB~A2Tune~\cite{ATL-PHYS-PUB-2012-003}+MSTW2008LO
			\item \pythia8+ATLAS~AZTune~\cite{ATLASZPt2}+CT14HERA2NNLO
			\item \pythia8+Tune2C~\cite{CDFTune}+CTEQ6L1
			\item \pythia8+Tune2M~\cite{CDFTune}+MRST~LO
			\item \pythia8+CMS~UE~Tune~CUETP8S1-CTEQ6L1~\cite{CMSTune}\\+CTEQ6L1
			\newline}
	\end{enumerate}
	
	A fast parametrized MC simulation (\pmcs), which is also used in our $W$ boson mass measurement~\cite{WMass,D0NewW}, is used to simulate electron identification efficiencies and the energy responses and resolutions of the electron and recoil system. 
	The \pmcs\ parameters are determined using a combination of GEANT3-based detailed simulation~\cite{geant} and control data samples. The primary control sample used for both the electromagnetic and hadronic response tuning is $Z\to ee$ events. Events recorded in random beam crossings are overlaid on $W$ and $Z$ boson events in the simulation to emulate the effect of additional collisions in the same or nearby beam bunch crossings.
	
	~\\
	\begin{center}
		\bf IV. DETECTOR RESPONSE CALIBRATION
	\end{center}
	~\\
	
	The $Z$ boson mass and width are used to calibrate the electromagnetic calorimeter energy response assuming a form $E^{\text{meas}} = \alpha\,E^{\text{true}} + \beta$, with constants $\alpha$ and $\beta$ determined from fits to the dielectron mass spectrum and the energy and angular distributions of the two electrons. The hadronic energy in the event contains the hadronic system recoiling from the $W$ boson, the effects of low energy products from spectator parton collisions and other beam collisions, final state radiation, and energy from the recoil particles that enters the electron selection window.
	The hadronic response (resolution) is calibrated using the mean (width) of the $\eta_{\text{imb}}$ distribution in $Z\to ee$ events in bins of the dielectron transverse momentum ($p_T^{ee}$). 
	Here, $\eta_{\text{imb}}$ is defined as the projection of the sum of $\vec{p}_T^{~ee}$ and $\vec{u}_T$ vectors on the axis bisecting the electron directions in the transverse plane~\cite{ua2eta}. More details can be found in Ref.~\cite{D0NewW}.
	
	
	~\\
	\begin{center}
		\bf V. BACKGROUNDS AND DATA/MC COMPARISONS
	\end{center}
	~\\

	The background in the $W$ boson candidate sample includes $Z\to ee$ events where one electron escapes detection, multijet events where a jet is misidentified as an electron with~~$\met$ arising from instrumental effects, and $W\to \tau\nu \to e\nu\nu\nu$ events. The $Z \to ee$ and multijet backgrounds are estimated from collider data, and the $W \to \tau \nu \to e \nu \nu \nu$ background is obtained from the \pmcs simulation of the process, as detailed in Ref.~\cite{D0NewW}. The fractions of these backgrounds relative to the signal are $1.08\%\pm 0.02$\% for $Z\to ee$, $1.02\%\pm 0.06$\% for multijet events, and $1.668\%\pm0.004$\% for $W\to\tau\nu\to e\nu\nu\nu$.
	
	\begin{figure*}[!]
		\begin{center}
			\begin{minipage}{1.0\textwidth}
				\resizebox{0.43\textwidth}{!}{\includegraphics{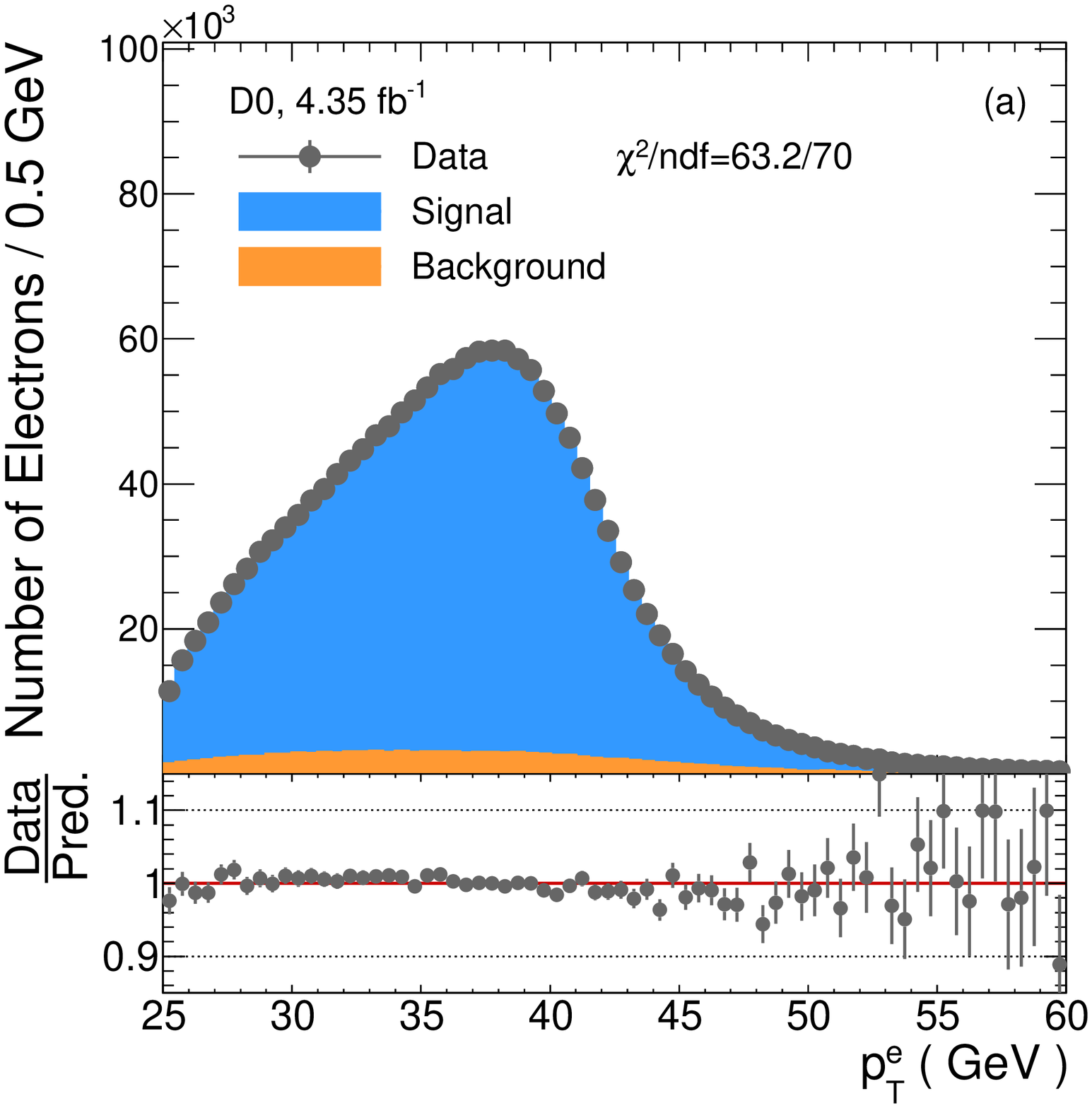}}
				\resizebox{0.43\textwidth}{!}{\includegraphics{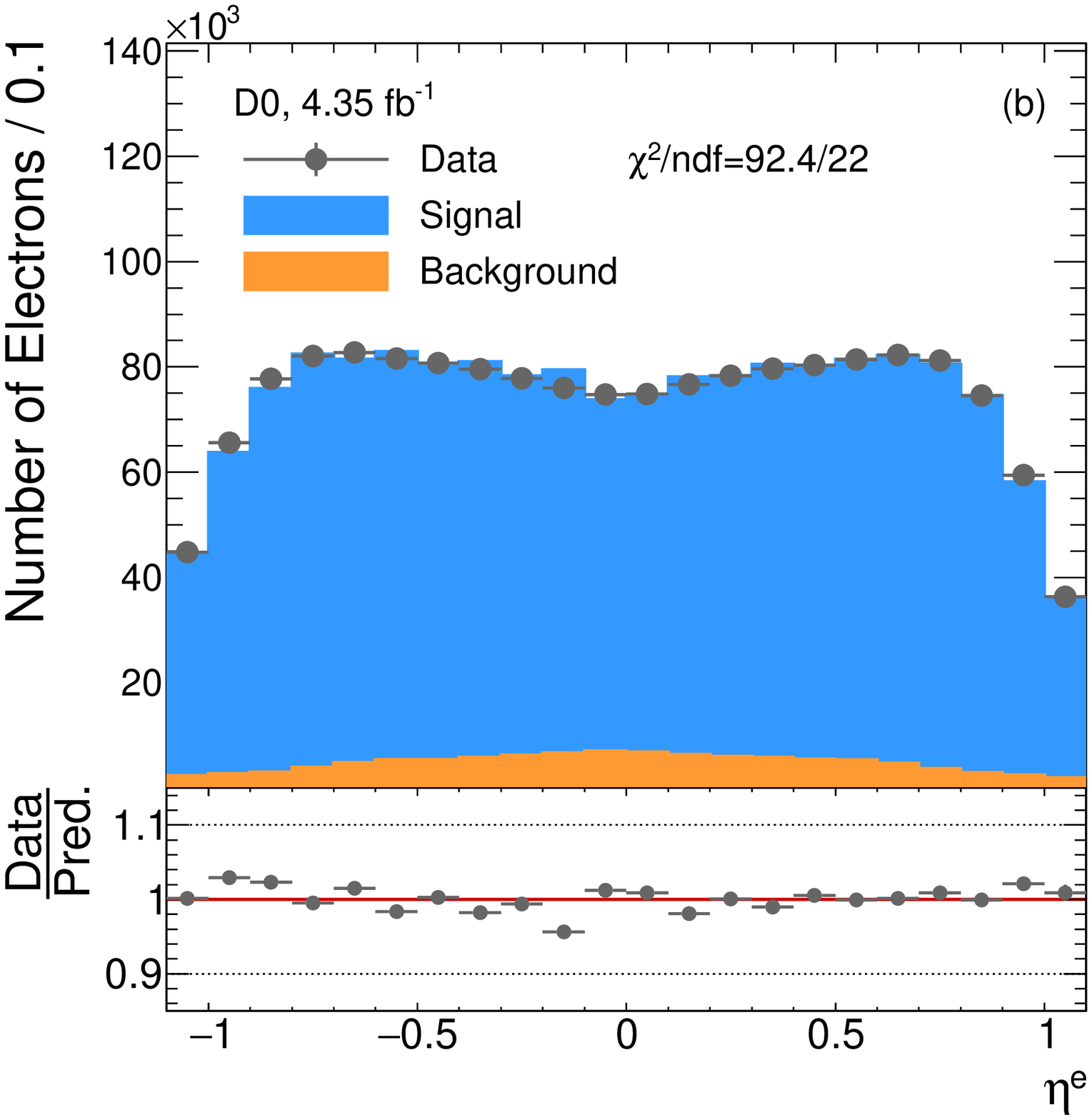}}
				\resizebox{0.43\textwidth}{!}{\includegraphics{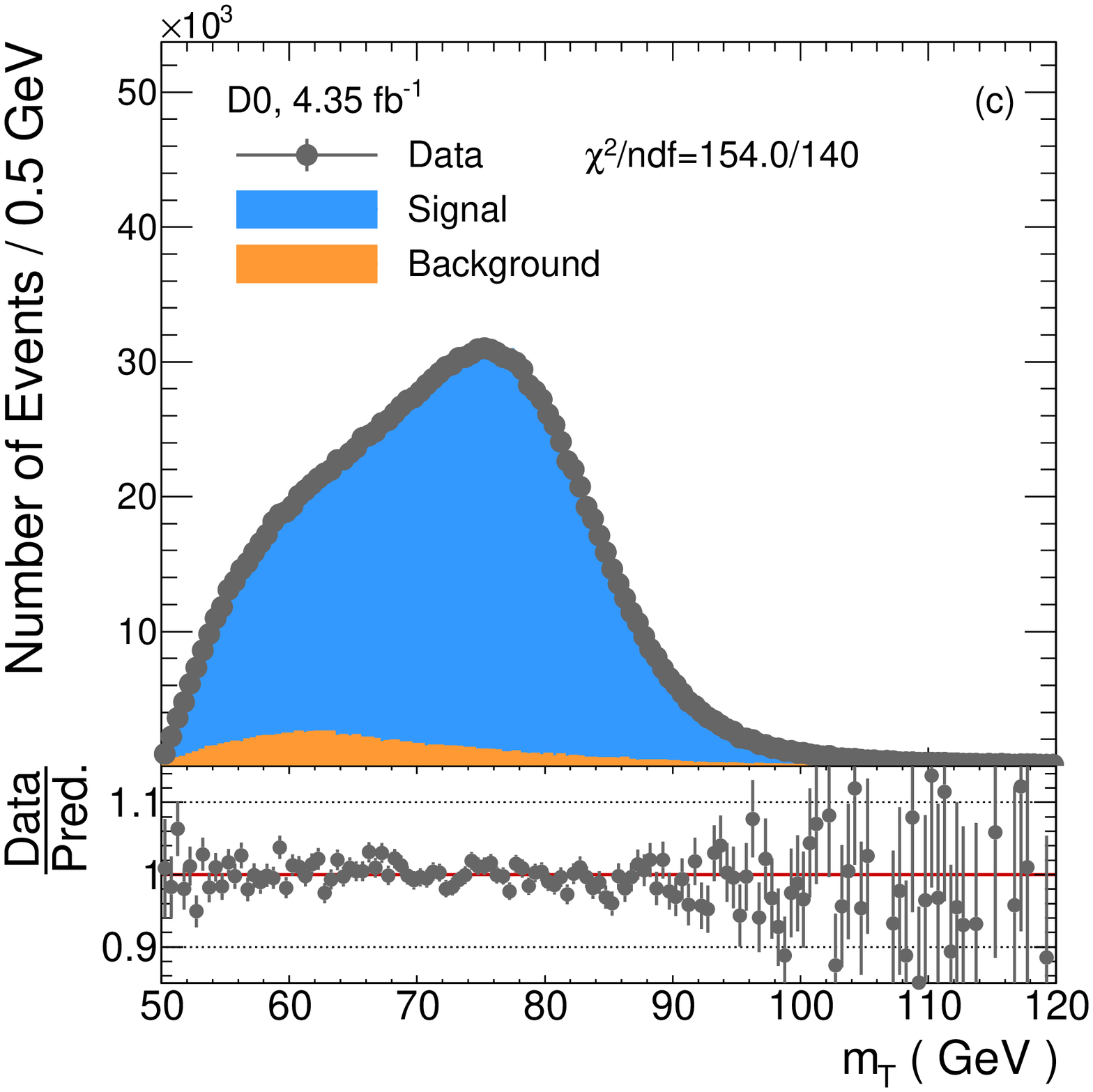}}
				\resizebox{0.43\textwidth}{!}{\includegraphics{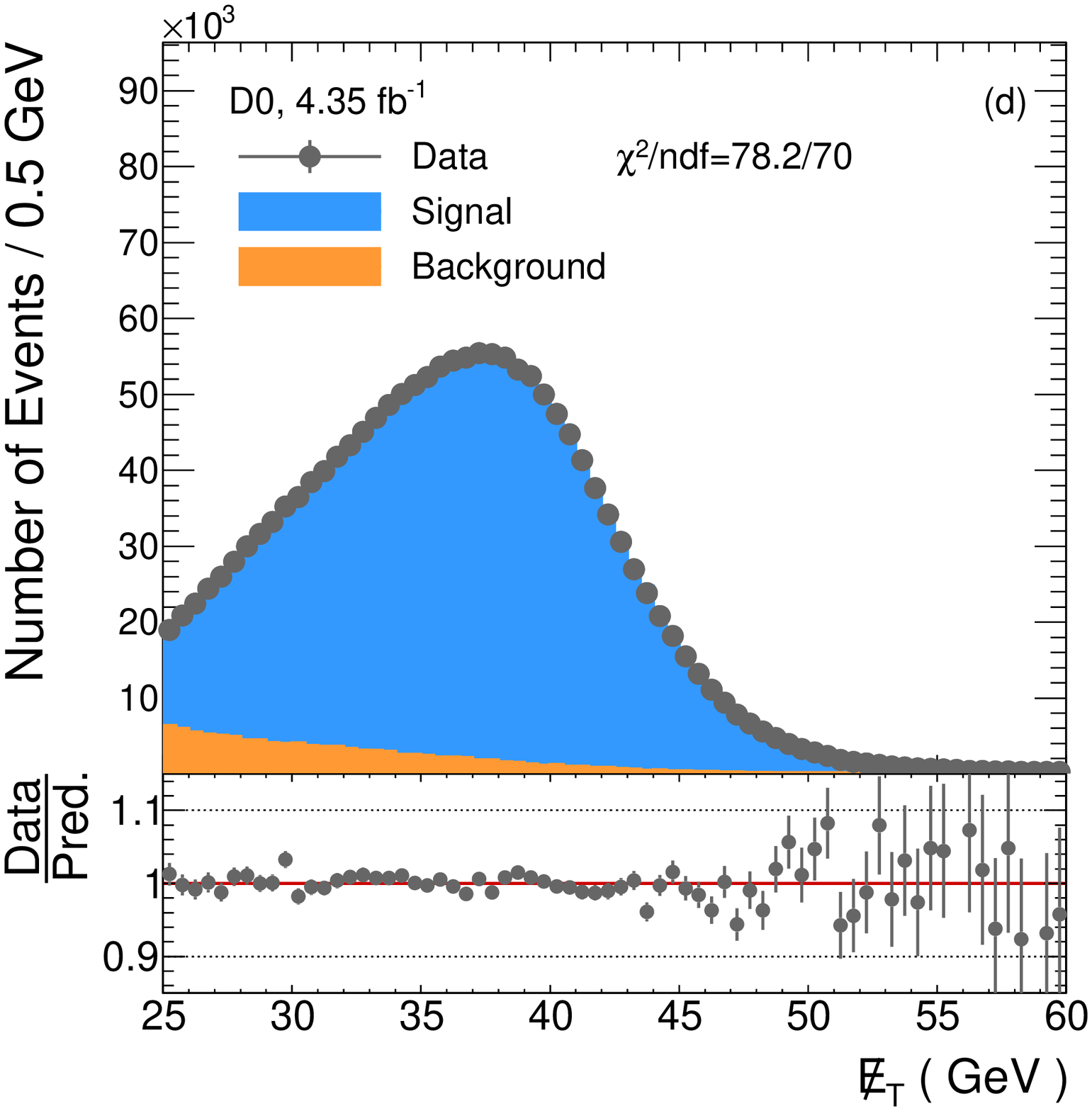}}
			\end{minipage}
			\caption{[color online] Kinematic distributions for (a) $p_{T}^{e}$, (b) $\eta^{e}$, (c) $m_{T}$, (d) $\met$. The data are compared to the PMCS plus background prediction in the upper panel, and the ratio of the data to the PMCS plus background prediction is shown in the lower panels. Only the statistical uncertainty is included.}
			\label{fig:DataMC}
		\end{center}
	\end{figure*}
	
	\begin{figure*}[!]
		\begin{center}
			\begin{minipage}{0.43\textwidth}
				\resizebox{1.0\textwidth}{!}{\includegraphics{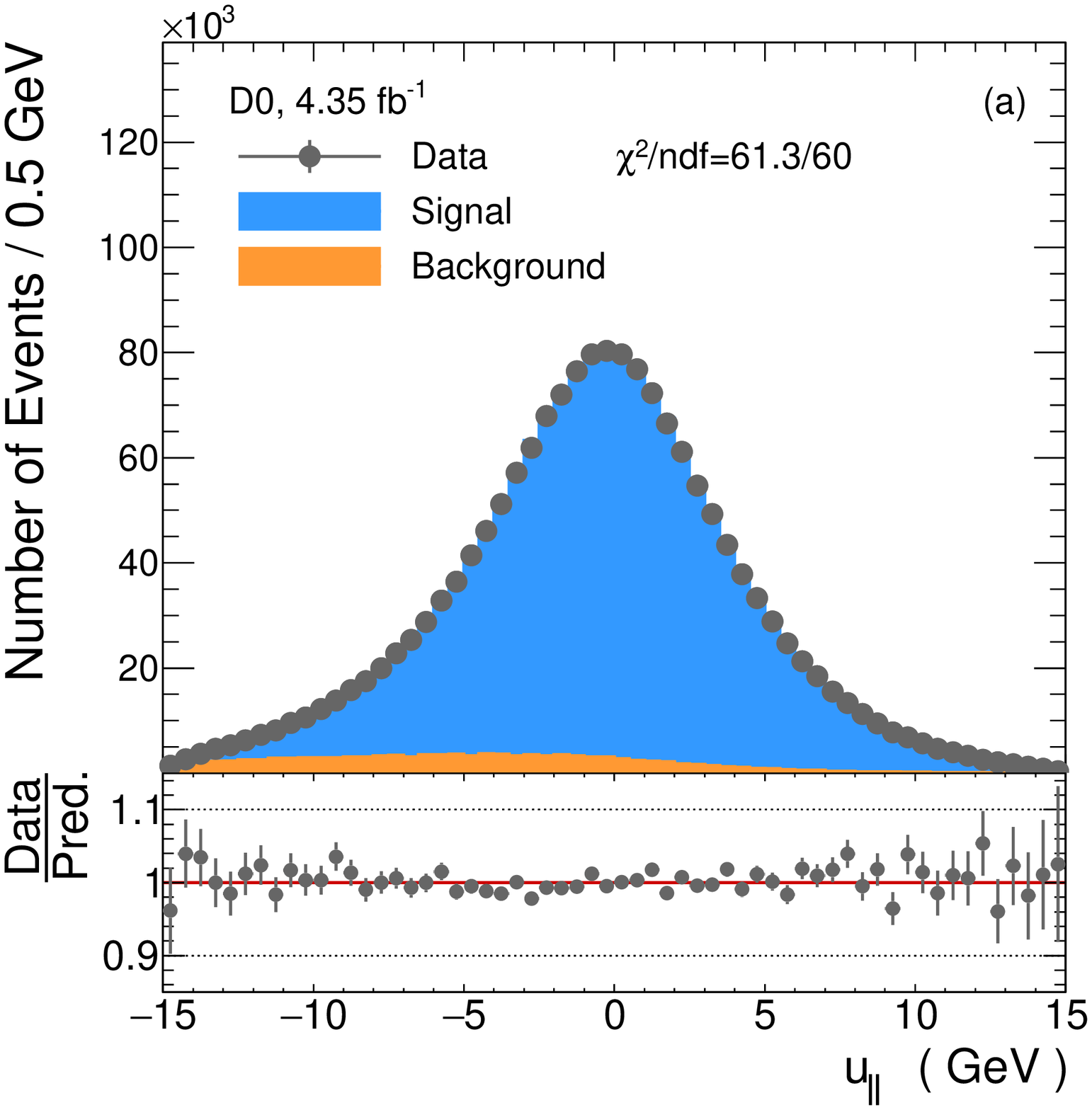}}
			\end{minipage}
			\begin{minipage}{0.43\textwidth}
				\resizebox{1.0\textwidth}{!}{\includegraphics{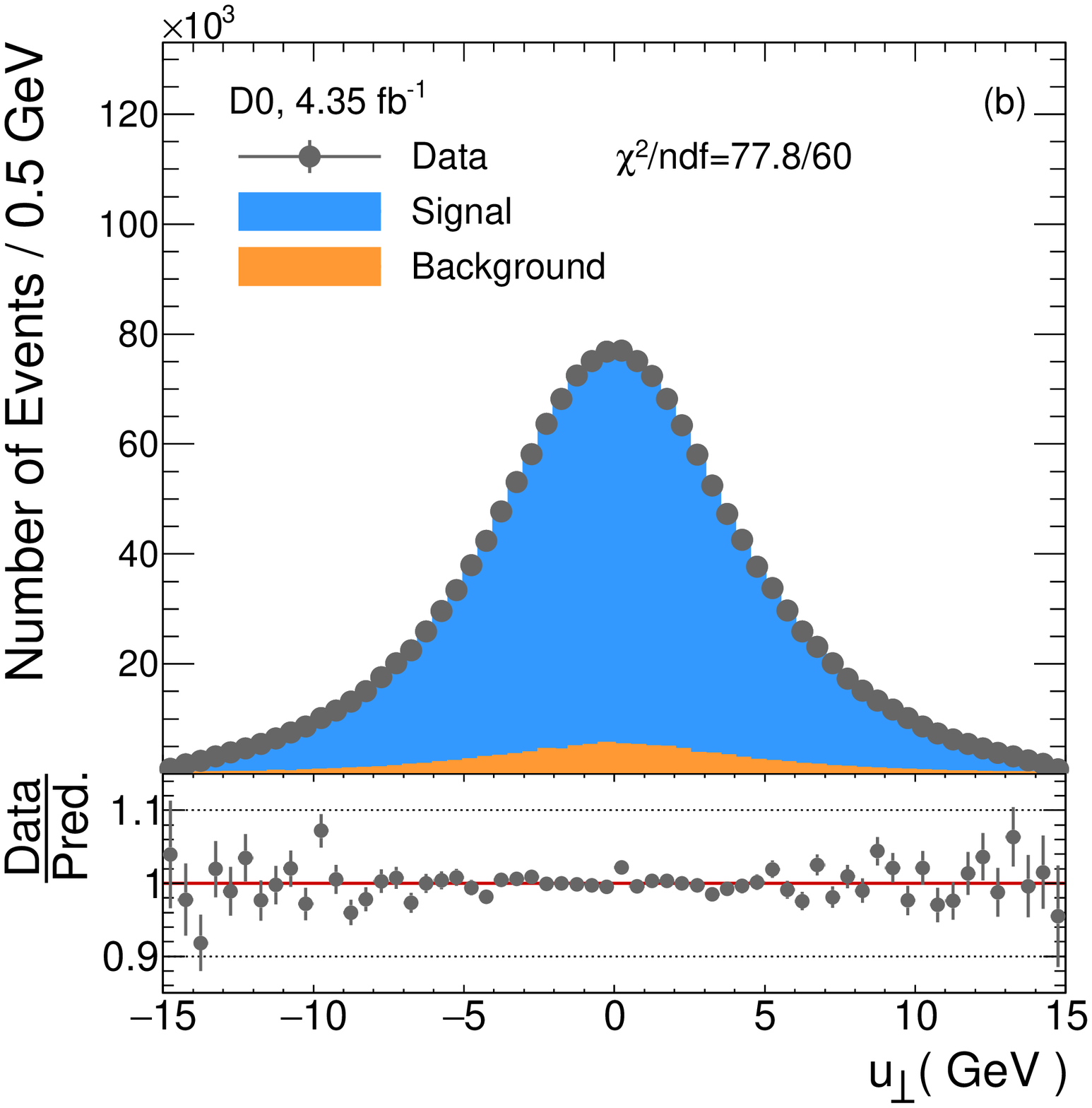}}
			\end{minipage}
			\caption{[color online] Kinematic distributions for (a) $u_{\parallel}$, (b) $u_{\perp}$. The data are compared to the PMCS plus background prediction in the upper panel, and the ratio of the data to the PMCS plus background prediction is shown in the lower panels. Only the statistical uncertainty is included.}
			\label{fig:DataMCRecoil}
		\end{center}
	\end{figure*}
	
	Several kinematic distributions of the signal predictions of \pmcs\ together with the expected background contributions taken from Ref.~\cite{D0NewW} are compared to data for $W$ boson candidate events in Figs.~\ref{fig:DataMC} and \ref{fig:DataMCRecoil}. The lepton transverse momentum, the lepton rapidity, the transverse mass, and the missing transverse energy shown in Fig.~\ref{fig:DataMC}, are not directly sensitive to $\pTW$ and therefore probe the general consistency of the simulation. To test the hadronic recoil modeling, we show in Fig.~\ref{fig:DataMCRecoil} the data and MC comparisons for the components of the hadronic recoil parallel to ($u_{\parallel}$) and perpendicular to ($u_{\perp}$) the direction of the electron. For all distributions in Figs.~\ref{fig:DataMC} and \ref{fig:DataMCRecoil}, the simulation is found to agree with the data.
	
	~\\
	\begin{center}
		\bf VI. ANALYSIS STRATEGY
	\end{center}
	~\\
	 
	The comparison of several $p_T^W$ models to data can be achieved either by comparing unfolded data directly with the predictions or by comparing predictions after accounting for detector response and resolution effects with background-subtracted data. Here folding refers to the modification of the model due to detector effects so as to compare directly to the reconstructed level data. Unfolding is the reverse transformation of the data to the particle level for comparison with the theoretical model. The limited $u_T$ detector resolution implies a large sensitivity to statistical fluctuations when unfolding, which have to be mitigated by a regularization scheme that increases the possible bias and thus the overall uncertainty. We therefore choose to perform the comparisons with the theory prediction at the reconstruction level.
	
	The folding of the different theory predictions with the \dzero detector response is based on the \pmcs~framework. In the first step, the baseline model of the $W$ boson production is reweighted in two dimensions, $p_T^W$ and $y^W$, to an alternative theory prediction to be tested. Here $y^W$ is the rapidity of the $W$ boson, which is highly correlated with $p_T^W$. In the second step, the reweighted theory prediction is used as input for the \pmcs~framework, resulting in detector level distributions of all relevant observables. In the third step, the uncertainties due to limited MC statistics, the hadronic recoil calibration, the electron identification and reconstruction efficiencies, as well as the electron energy response are estimated for each theory prediction by varying all relevant detector response parameters of the \pmcs~framework within their uncertainties. Uncertainties due to limited MC statistics, the uncertainties due to the electron identification and reconstruction efficiencies as well as the electron energy response are found to be negligible for the $u_T$ distribution. The hadronic recoil calibration is modeled by five calibration parameters~\cite{D0NewW}. These five parameters are divided into two groups, one containing three parameters for the response of $u_T$ and the other containing two parameters for the resolution of $u_T$. Only the parameters in the same group are considered to be correlated. Given the correlation matrices of these two groups of parameters, these five parameters are transformed into another five uncorrelated parameters by a linear combination. Each component of the hadronic recoil uncertainty is estimated by varying one of the five uncorrelated parameters with its uncertainty. The combined hadronic recoil uncertainty is calculated by adding in quadrature the individual components in each $u_T$ bin. The uncertainty from each component is considered to be bin-by-bin correlated, and the uncertainties from different components are assumed to be uncorrelated. 
	
	The uncertainties on the measured $u_T$ distribution of the background-subtracted data are the statistical uncertainty, which is treated as bin-to-bin uncorrelated, and the uncertainty due to the background, which is significantly smaller than the statistical uncertainty. The background uncertainty is obtained by varying the overall number of events from each background contribution independently within its uncertainty, so this uncertainty should be considered to be bin-by-bin correlated. Because the uncertainties are small, the effects of these correlations are found to be negligible.
	
	The resulting fractions of events in the $u_T$ bins with boundaries $[0,2,5,8,11,15]~\text{GeV}$ are summarized in Table \ref{tab:result_data} for the background-subtracted data along with the combined statistical and systematic uncertainties.
	
	
	\begin{table*}[!]
		\begin{center}
			\caption{\small The fraction of $W$ boson events in bins of $u_T$ for the background-subtracted data. The combined statistical and systematic uncertainties are shown.}
			\label{tab:result_data}
			\begin{tabular}{l c c c c c }
				\hline
				$u_T$ bin	& 0--2 GeV		& 2--5 GeV		& 5--8 GeV		& 8--11 GeV	& 11--15 GeV	\\
				\hline
				Fraction of events in the $u_T$ bin	& 0.1181 & 0.3603& 0.2738 & 0.1515 & 0.0963 \\
				Total uncertainty 		& 0.0003 & 0.0005 & 0.0005 & 0.0004 & 0.0003 \\
				\hline
			\end{tabular}
			
		\end{center}
	\end{table*}
	
\begin{figure*}[!]
   	\begin{center}
   		\includegraphics[width=0.43\textwidth]{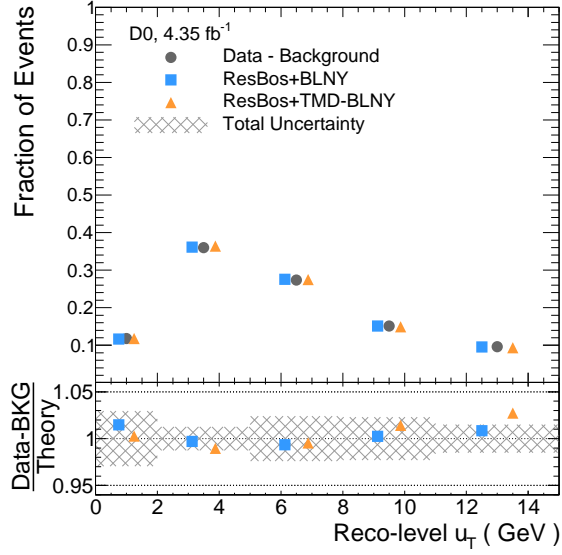}
   		\caption{[color online] \small Comparisons of the measured and predicted $u_T$ distributions after the detector response simulation for different MC predictions based on \resbos. The ratios of the background-subtracted data to each theory prediction are shown in the lower panel together with the 1$\sigma$ uncertainty band. The total experimental uncertainty is indicated by the hatched band; it is dominated by the uncertainty due to the hadronic recoil calibration. The points for the predictions are offset horizontally to aid with visibility.}
   		\label{fig:ComparisonResBos}
   	\end{center}
   \end{figure*}

\begin{figure*}[!]
   	\begin{center}
   		\includegraphics[width=0.43\textwidth]{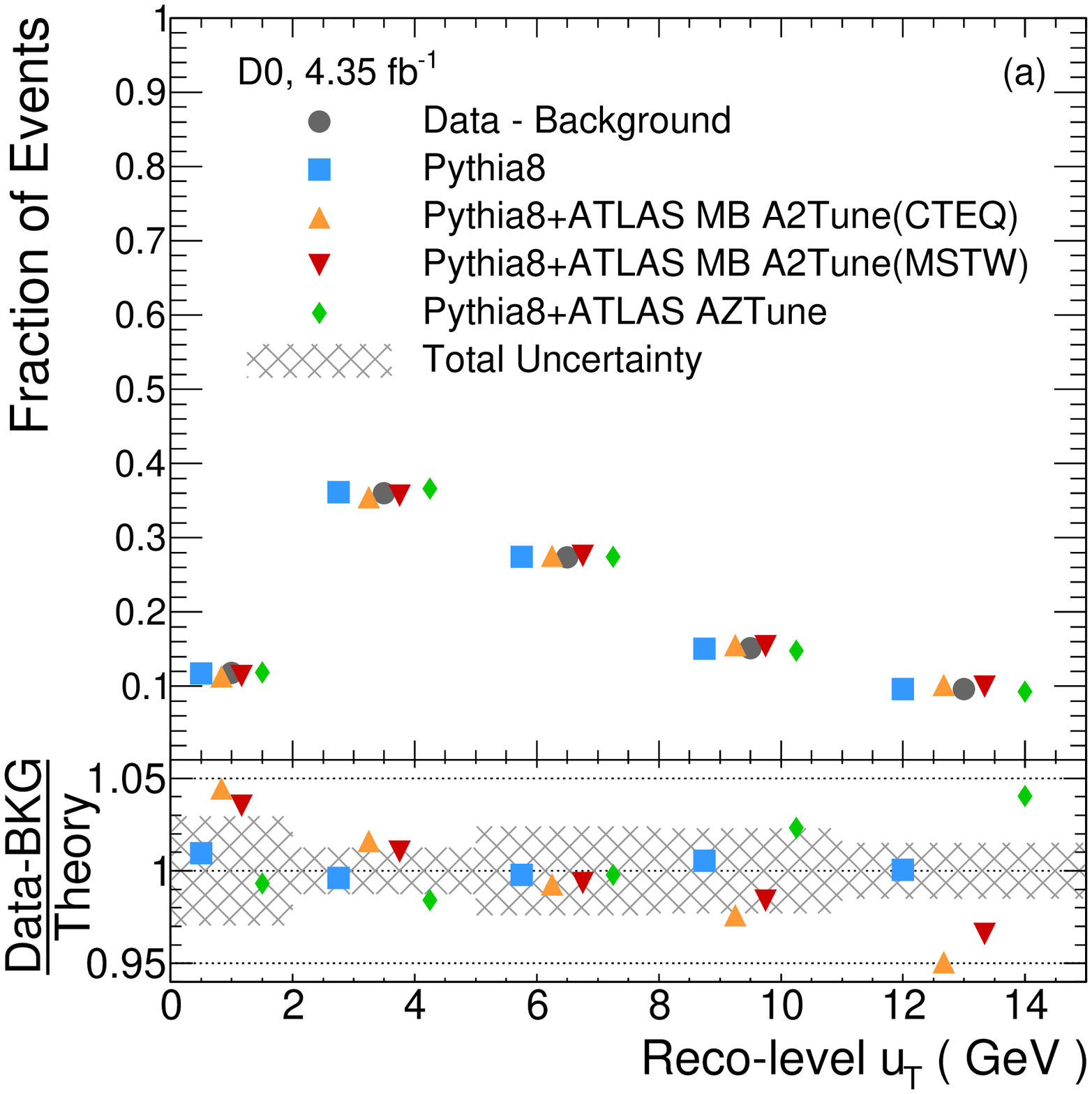}
   		\includegraphics[width=0.43\textwidth]{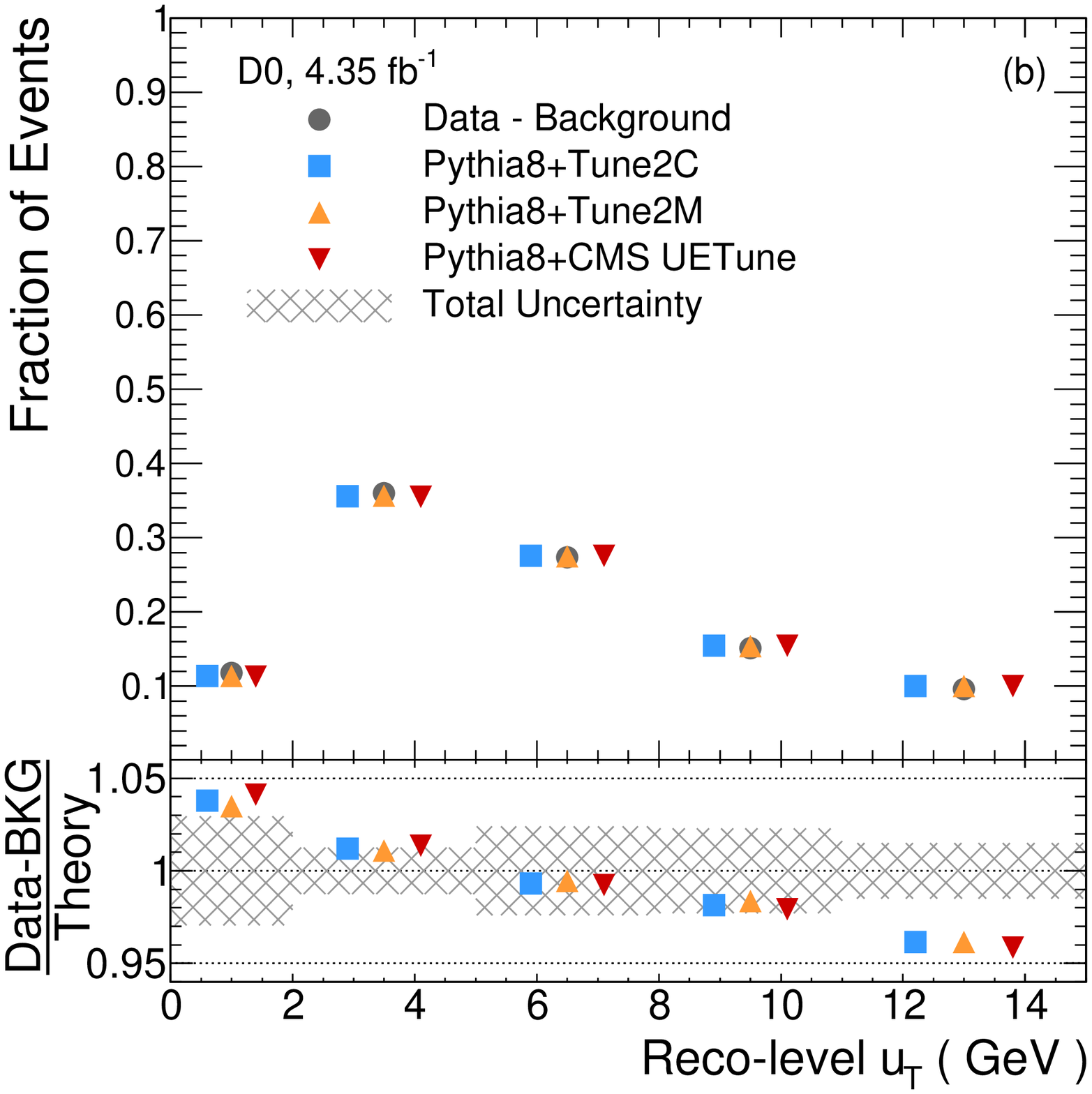}
   		\caption{[color online] \small Comparisons of the measured and predicted $u_T$ distributions after the detector response simulation for different MC predictions based on \pythia. The ratios of the background-subtracted data to each theory prediction are shown in the lower panel together with the 1$\sigma$ uncertainty band. The total experimental uncertainty is indicated by the hatched band; it is dominated by the uncertainty due to the hadronic recoil calibration. The points for the predictions are offset horizontally to aid with visibility.}
   		\label{fig:ComparisonPythia}
   	\end{center}
   \end{figure*}

	~\\
	\begin{center}
		\bf VIII. RESULTS AND COMPARISON TO THEORY
	\end{center}
	~\\
   At the reconstruction level, the $u_T$ distribution of the background-subtracted data is compared to the predictions of \resbos and \pythia listed in Section III. The predictions are normalized to the background-subtracted data with $u_T<15$~GeV. The data are compared to \resbos predictions based on two different non-perturbative functions, BLNY and TMD-BLNY in Fig.~\ref{fig:ComparisonResBos}. Figure \ref{fig:ComparisonPythia} shows comparisons with \pythia predictions using the different tunes provided by several collaborations. All five $u_T$ bins are considered in the $\chi^2$ calculation. The uncertainties due to the resummation calculation of \resbos and the tune of \pythia are not considered in the comparison and the $\chi^2$ calculation, and the uncertainty due to the PDF set is negligible. Since both the data and the prediction are normalized to unity, the number of degrees of freedom is 4. The resulting $\chi^2/$ndf values for all models and the corresponding significances in the Gaussian approximation are summarized in Table \ref{tab:result_chi2}. From this comparison, \pythia8+ATLAS MB A2Tune+CTEQ6L1 is excluded with a $p$-value equal to $5.84\times10^{-10}$ and \pythia8+CMS UE Tune CUETP8S1-CTEQ6L1+CTEQ6L1 is excluded with a $p$-value equal to $4.23\times10^{-7}$. All the other \pythia8 predictions except the default, \pythia8+CT14HERA2NNLO, are disfavored. The model based on {\sc resbos}+BLNY agrees with the data.
   
   \begin{table*}[!]
   	\begin{center}
   		\caption{\small Chi-squared per degree of freedom and the corresponding $p$-value for the reconstructed-level comparison. Significance is the number of standard deviations in the Gaussian approximation for the difference between each model and the background-subtracted data. Since the distributions are normalized to unity before the comparison, the number of degrees of freedom is 4.}
   		\label{tab:result_chi2}
   		\begin{tabular}{l c c c}
   			\hline
   			Generator/Model	\hspace{5.5cm}		& $\chi^2/$ndf	& $p$-value & Signif.\\
   			\hline
   			\resbos(Version CP 020811)+BLNY+CTEQ6.6	& 0.49 & $7.41\times10^{-1}$& 0.33				\\
   			\resbos(Version CP 112216)+TMD-BLNY+CT14HERA2NNLO & 3.13 & $1.39\times10^{-2}$ & 2.46				\\
   			\pythia8+CT14HERA2NNLO	& 0.32 & $8.63\times10^{-1}$ & 0.17					\\
   			\pythia8+ATLAS MB A2Tune+CTEQ6L1	& 12.25 & $5.84\times10^{-10}$ & 6.19					\\
   			\pythia8+ATLAS MB A2Tune+MSTW2008LO	& 6.17 & $5.83\times10^{-5}$ & 4.02				\\
   			\pythia8+ATLAS AZTune+CT14HERA2NNLO	& 6.61 & $2.60\times10^{-5}$ & 4.21				\\
   			\pythia8+Tune2C+CTEQ6L1 & 7.66 & $3.61\times10^{-6}$ & 4.63\\
   			\pythia8+Tune2M+MRSTLO & 7.32 & $6.89\times10^{-6}$ & 4.50\\
   			\pythia8+CMS UE Tune CUETP8S1-CTEQ6L1+CTEQ6L1 & 8.80 & $4.23\times10^{-7}$ & 5.06\\
   			\hline 
   		\end{tabular}
   		
   	\end{center}
   \end{table*}

	~\\
	\begin{center}
		\bf IX. CONCLUSION
	\end{center}
	~\\
	We report a study of the normalized transverse momentum distribution of $W$ bosons produced in $p\bar{p}$ collisions at a center of mass energy of 1.96 TeV, using 4.35 fb$^{-1}$ of data collected by the \dzero collaboration at the Fermilab Tevatron collider. The $u_T$ distribution of the data is compared to those from several theory predictions at the reconstruction level. From these comparisons, \pythia8+ATLAS MB A2Tune+CTEQ6L1 and \pythia8+CMS UE Tune CUETP8S1- CTEQ6L1+CTEQ6L1 are excluded. All the other \pythia8 predictions except the default, \pythia8+CT14HERA2NNLO, are disfavored. Both models based on \resbos give satisfactory fits to the data. The precision is limited by the uncertainty due to the hadronic recoil calibration. 
	
	In the appendix, we describe a procedure by which theoretical models for the $p_T$ distribution of $W$ boson production beyond those considered in this paper can be quantitatively compared to the D0 data.
	
	This study is the first inclusive $\pTW$ analysis using Tevatron Run II data. Our data are binned sufficiently finely in $\pTW$ to resolve the peak in the cross section, unlike the previous measurements at the LHC. In comparison to measurements by LHC experiments, which involve sea quarks, this work provides additional information for evaluating resummation calculations of transverse momentum of $W$ bosons when the production is dominated by valence quarks.

	~\\
	~\\
	\begin{center}
		\bf ACKNOWLEDGEMENTS
	\end{center}
	~\\
	%
	
	This document was prepared by the D0 collaboration using the resources of the Fermi National Accelerator Laboratory (Fermilab),
	a U.S. Department of Energy, Office of Science, HEP User Facility. Fermilab is managed by Fermi Research Alliance, LLC (FRA),
	acting under Contract No. DE-AC02-07CH11359.
	
	We thank the staffs at Fermilab and collaborating institutions,
	and acknowledge support from the
	Department of Energy and National Science Foundation (United States of America);
	Alternative Energies and Atomic Energy Commission and
	National Center for Scientific Research/National Institute of Nuclear and Particle Physics  (France);
	Ministry of Education and Science of the Russian Federation, 
	National Research Center ``Kurchatov Institute" of the Russian Federation, and 
	Russian Foundation for Basic Research  (Russia);
	National Council for the Development of Science and Technology and
	Carlos Chagas Filho Foundation for the Support of Research in the State of Rio de Janeiro (Brazil);
	Department of Atomic Energy and Department of Science and Technology (India);
	Administrative Department of Science, Technology and Innovation (Colombia);
	National Council of Science and Technology (Mexico);
	National Research Foundation of Korea (Korea);
	Foundation for Fundamental Research on Matter (The Netherlands);
	Science and Technology Facilities Council and The Royal Society (United Kingdom);
	Ministry of Education, Youth and Sports (Czech Republic);
	Bundesministerium f\"{u}r Bildung und Forschung (Federal Ministry of Education and Research) and 
	Deutsche Forschungsgemeinschaft (German Research Foundation) (Germany);
	Science Foundation Ireland (Ireland);
	Swedish Research Council (Sweden);
	China Academy of Sciences and National Natural Science Foundation of China (China);
	and
	Ministry of Education and Science of Ukraine (Ukraine).
	%

    ~\\
    \begin{center}
    	\bf APPENDIX. DETECTOR RESPONSE FOR FUTURE COMPARISONS
    \end{center}
    ~\\
    
    In order to compare additional model predictions to the measured data, some previous measurements~\cite{DZeroWPt2, ATLASWPt, CMSPt} have been unfolded to the particle level.
    However, in this study, instead of providing the unfolded particle level $\pTW$ distribution, a fast folding procedure is introduced for two reasons: first, no new piece of information would be added by the unfolding procedure so the precision on the particle level would not be better than that on the reconstruction level. Due to the systematic uncertainty from the MC modeling or the regularization which would be introduced by an unfolding method, the precision of the unfolded particle level distribution would be reduced. This reduction would be greater when the resolution of the distribution is worse, and it would be smaller when the bin width is enlarged. But when the bin width is too large, the rise and hence the shape of the spectrum cannot be resolved. Second, it is hard to estimate the bin-by-bin correlation of the uncertainty due to the MC modeling or the regularization properly, since the definitions of these uncertainties are often arbitrary. Therefore, the folding method provided gives a more precise and reliable means of comparison than would an unfolded result. 
 
    This fast folding procedure has to be applied on $p_T^W$ spectra within the fiducial region defined by an electron with $p_{T}^{e}>25$ GeV and $|\eta^{e}|<1.05$, a $W$ boson with $50 < m_{T}<200$ GeV and a neutrino with $p_{T}^{\nu}>25$ GeV. The numbers of events in $p_T^W$ bins with boundaries $[0,2,5,8,11,15,600]~\text{GeV}$ are the input to this folding procedure.
    
    In the first step, the spectrum has to be corrected for the detector efficiency in each $p_T^W$ bin, via
    \[X_i^{{\mathrm{corr}}} = {\cal{E}}_{i} X_{i}.\]
    Here $X_i$ is the number of events in bin $i$ of the $p_T^W$ distribution within the fiducial region, ${\cal{E}}_i$ is the detector efficiency summarized in Table \ref{tab:Efficiency} and $X_i^{{\mathrm{corr}}}$ is the number of efficiency-corrected events on the particle level in bin $i$. Even though most of the events with $p_T^W>100$~GeV will not satisfy $u_T<15$~GeV after the \pmcs simulation, we still chose $600$~GeV as the upper edge of the last $\pTW$ bin. This is because the efficiency correction in the last $\pTW$ bin is directly related to this choice, and the upper edge of the last $\pTW$ bin should be kept the same as the value used when deriving those efficiency correction factors.
   
    \begin{table*}[!]
    	\begin{center}
    		\caption{\small The efficiency correction ${\cal{E}}(p_T^W)$ in each $p_T^W$ bin. The efficiency correction is the probability to pass theion selection for the events that pass the particle level selection.}
    		\label{tab:Efficiency}
    		\begin{tabular}{lcccccc}
    			\hline
    			$p_T^W$ bin	& 0--2 GeV		& 2--5 GeV		& 5--8 GeV		& 8--11 GeV	& 11--15 GeV	& 15--600 GeV	\\
    		   	\hline
    			${\cal{E}}(p_T^W)$ & 0.2330 & 0.2367& 0.2387 & 0.2396 & 0.2385 & 0.2332\\
    			\hline
    		\end{tabular}
    		
    	\end{center}
    \end{table*}

   The second step accounts for the mapping from $p_T^W$ to $u_T$ using the response matrix $R_{ij}$ via
    \[N_i = \sum_{j=1}^{6} R_{ij} X_j^{{\mathrm{corr}}},\]
    where $N_i$ is the resulting number of events of the reconstruction level in bin $i$ and $R_{ij}$ is a $5\times6$ matrix. The response matrix is obtained for the signal sample using the \pmcs framework and it is summarized in Table \ref{tab:Matrix}.
    
    \begin{table*}[!]
    	\begin{center}
    		\caption{\small Detector response matrix. The number in each cell is the probability for the events in one $\pTW$ bin to be reconstructed into different $u_T$ bins.}
    		\label{tab:Matrix}
    		\begin{tabular}{lcccccc}
    			\hline
    			$p_T^W$ bin	& 0--2 GeV		& 2--5 GeV		& 5--8 GeV		& 8--11 GeV	& 11--15 GeV	& 15--600 GeV	\\
    			\hline
    			$0<u_T<2$ GeV & 0.1784 & 0.1696& 0.1212 & 0.0745 & 0.0372 & 0.0069\\
    			$2<u_T<5$ GeV & 0.4636 & 0.4588& 0.4109 & 0.3163 & 0.1974 & 0.0452\\
    			$5<u_T<8$ GeV & 0.2452 & 0.2524& 0.2966 & 0.3331 & 0.3146 & 0.1121\\
    			$8<u_T<11$ GeV & 0.0806 & 0.0863& 0.1193 & 0.1810 & 0.2495 & 0.1637\\
    			$11<u_T<15$ GeV & 0.0269 & 0.0270& 0.0428 & 0.0775 & 0.1550 & 0.2210\\
    			\hline
    		\end{tabular}
    		
    	\end{center}
    \end{table*}
    
    In the third step, after the application of the response matrix, the resulting spectrum has to be corrected for events which would have passed the reconstruction level cuts but not the particle level selection, via
    \[N_i^{{\mathrm{corr}}} = \frac{N_i}{F_i}.\]
    Here $F_i$ is the fiducial correction factor in $u_T$ bin $i$ and $N_i^{{\mathrm{corr}}}$ is the number of fiducial-corrected events on the reconstruction level in bin $i$.
    The corresponding fiducial correction factors are derived from the nominal signal sample using \pmcs and are summarized in Table \ref{tab:Fiducial}. 
    
    \begin{table*}[!]
    	\begin{center}
    		\caption{\small The fiducial correction $F(u_T)$ in each $u_T$ bin. The fiducial correction is the probability to pass the particle level selection for the events that pass theion selection.}
    		\label{tab:Fiducial}
    		\begin{tabular}{lccccc}
    			\hline
    			$u_T$ bin	& 0--2 GeV		& 2--5 GeV		& 5--8 GeV		& 8--11 GeV	& 11--15 GeV		\\
    			\hline
    			$F(u_T)$ & 0.8624 & 0.8689& 0.8797 & 0.8812 & 0.9036 \\
    			\hline
    		\end{tabular}
    		
    	\end{center}
    \end{table*}
    
    Finally, in order to get the shape of the distribution, the folded $u_T$ distribution is normalized to unity.  The fraction of the events in each $u_T$ bin, ${\cal{N}}_i$, is calculated via the following formula:
    \[{\cal{N}}_i = \frac{N_i^{{\mathrm{corr}}}}{\sum_{j=1}^{5} N_j^{{\mathrm{corr}}}}\]
    This normalized $u_T$ distribution is the folded result, which can be compared to the background-subtracted data directly.
    
    This fast folding procedure is demonstrated to give reconstruction level distributions consistent with those provided by \pmcs for the models studied in this paper. Both the efficiency correction and the response matrix are applied directly to the $\pTW$ distribution and hence no model assumptions are made. However, the fiducial correction could depend on details of the theoretical model used. We have tested this possibility using two toy production models which differ from our baseline model by either shifting the peak in the $\pTW$ distribution by 20\% or by broadening the peak by about 20\%.  In these cases, the $u_T$ distributions resulting from the fast folding procedure differed negligibly from those using \pmcs. 
    
    In order to calculate the chi-square value for the difference between the folded theory prediction and the background-subtracted data, the uncertainty of the folded distribution in each $u_T$ bin and the bin-by-bin correlation matrix are also needed. In this fast folding procedure, the detector response is represented by two corrections, the fiducial correction and the efficiency correction, and one detector response matrix. Since the systematic uncertainty is estimated from the difference in the normalized $u_T$ distribution between the nominal response and the systematic variation, the uncertainty and the correlation matrix are model dependent, which is why the folding inputs for all of the systematic variations must be provided. 
    
    The uncertainty on the $u_T$ distribution consists of three independent parts: the uncertainty due to the MC statistics, the uncertainty due to the hadronic recoil calibration, and the uncertainty due to the electron identification and reconstruction efficiencies and the electron energy response. The dominant uncertainty is the one due to the hadronic recoil. The uncertainty due to the MC statistics is directly provided in Table \ref{tab:mc_stat_unc}, which is considered to be bin-by-bin uncorrelated. 
    
    \begin{table*}[!]
    	\begin{center}
    		\caption{\small The systematic uncertainty due to the MC statistics in each $u_T$ bin of the folded result.}
    		\label{tab:mc_stat_unc}
    		\begin{tabular}{l c c c c c }
    			\hline
    			$u_T$ bin	& 0--2 GeV		& 2--5 GeV		& 5--8 GeV		& 8--11 GeV	& 11--15 GeV	\\
    		    \hline
    			Uncertainty due to the MC statistics in the folded $u_T$ distribution 		& 0.0005 & 0.0007 & 0.0006 & 0.0005 & 0.0004 \\
    			\hline
    		\end{tabular}
    	\end{center}
    \end{table*}
    \begin{table*}[!]
    	\begin{center}
    		\caption{\small The efficiency correction ${\cal{E}}(p_T^W)$ in each $p_T^W$ bin from eleven systematic variations. The efficiency correction is the probability to pass the reconstruction level selection for the events that pass the particle level selection. The first ten systematic variations are for the uncertainty due to the hadronic recoil and the last one is for the uncertainty due to the electron energy response.}
    		\label{tab:syst_var_Efficiency}
    		\begin{tabular}{lcccccc}
    			\hline
    			$p_T^W$ bin	& 0--2 GeV		& 2--5 GeV		& 5--8 GeV		& 8--11 GeV	& 11--15 GeV	& 15--600 GeV	\\
    		    \hline
    			Systematic Variation No.~1 & 0.2348 & 0.2374 & 0.2377 & 0.2405 & 0.2392 & 0.2332 \\ 
    			Systematic Variation No.~2 & 0.2345 & 0.2370 & 0.2392 & 0.2377 & 0.2382 & 0.2334 \\
    			Systematic Variation No.~3 & 0.2336 & 0.2374 & 0.2388 & 0.2377 & 0.2378 & 0.2317 \\ 
    			Systematic Variation No.~4 & 0.2335 & 0.2369 & 0.2394 & 0.2385 & 0.2379 & 0.2329 \\ 
    			Systematic Variation No.~5 & 0.2323 & 0.2365 & 0.2392 & 0.2385 & 0.2393 & 0.2326 \\ 
    			Systematic Variation No.~6 & 0.2337 & 0.2355 & 0.2390 & 0.2408 & 0.2387 & 0.2321 \\ 
    			Systematic Variation No.~7 & 0.2342 & 0.2373 & 0.2384 & 0.2386 & 0.2390 & 0.2318 \\ 
    			Systematic Variation No.~8 & 0.2328 & 0.2362 & 0.2384 & 0.2386 & 0.2390 & 0.2322 \\ 
    			Systematic Variation No.~9 & 0.2360 & 0.2369 & 0.2382 & 0.2398 & 0.2376 & 0.2323 \\
    			Systematic Variation No.~10 & 0.2327 & 0.2371 & 0.2387 & 0.2390 & 0.2387 & 0.2328 \\
    			Systematic Variation No.~11 & 0.2343 & 0.2370 & 0.2379 & 0.2399 & 0.2374 & 0.2315 \\ 
    			\hline
    		\end{tabular}
    	\end{center}
    \end{table*}

	The other two parts of the uncertainty should be estimated with systematic variations. There are eleven systematic variations provided in total, ten for the uncertainty due to the hadronic recoil calibration and one for the uncertainty due to the efficiency and the energy response of the electron. The hadronic recoil response and resolution are characterized by the five uncorrelated parameters discussed in Section VI. The uncertainties due to positive and negative changes in these parameters differ, so we must evaluate both signs of parameter change, thus giving the first ten variations. The eleventh systematic variation is derived with the parameter $\alpha$, which is mentioned in Sec. IV, changed by its uncertainty. This is an overestimation of the uncertainty due to the strong anti-correlation between $\alpha$ and $\beta$. The folding inputs of these eleven systematic variations are provided in Tables \ref{tab:syst_var_Efficiency}, \ref{tab:syst_var_Matrix} and \ref{tab:syst_var_Fiducial}. The uncertainties from different variations are considered to be uncorrelated and the uncertainty from each variation is considered to be bin-by-bin correlated. The bin-by-bin covariance matrix of systematic variation $k$ is defined as $\Sigma^{(k)}$, whose element is calculated via
	\[\Sigma^{(k)}_{ij} = ({\cal{N}}_i - {\cal{N}}_i^{(k)})\times({\cal{N}}_j - {\cal{N}}_j^{(k)}).\]
	Here ${\cal{N}}_i^{(k)}$ is the folded result from systematic variation $k$. The covariance matrix of the uncertainty due to the hadronic recoil calibration are calcualted by the average of the covariance matrices from the positive and negative changes. The covariance matrix of the total systematic uncertainty, $\Sigma^{{\mathrm{(Syst.)}}}$, is calculated as the sum of the covariance matrix of the uncertainty due to the hadronic recoil calibration and that of the uncertainty due to the efficiency and the energy response of the electron, via
	\[\Sigma^{{\mathrm{(Syst.)}}} = \frac{\sum_{k=1}^{10} \Sigma^{(k)}}{2} + \Sigma^{(11)}.\]
	The total uncertainty of the folded result is the combination of the statistical uncertainty and the total systematic uncertainty.
	The total covariance matrix used in the $\chi^2$ calculation, $\Sigma^{{\mathrm{(Total)}}}$, is the sum of the covariance matrix of the systematic uncertainty and the statistical uncertainties due to both data and MC statistics, $\Sigma^{{\mathrm{(Data~stat.)}}}$ and $\Sigma^{{\mathrm{(MC~stat.)}}}$, via
	\[\Sigma^{{\mathrm{(Total)}}} = \Sigma^{{\mathrm{(Data~stat.)}}}+\Sigma^{{\mathrm{(MC~stat.)}}} + \Sigma^{{\mathrm{(Syst.)}}}.\]
	Here $\Sigma^{{\mathrm{(Data~stat.)}}}$ is a diagonal matrix constructed with the total uncertainty provided in Table \ref{tab:result_data} and $\Sigma^{{\mathrm{(MC~stat.)}}}$ is also a diagonal matrix constructed with the uncertainty summarized in Table \ref{tab:mc_stat_unc}.
	
    \begin{table*}[!]
    	\begin{center}
    		\caption{\small Detector response matrices for the eleven systematic variations. The numbers in each cell are the probability for the events in one $\pTW$ bin to be reconstructed into different $u_T$ bins. The first ten systematic variations are for the uncertainty due to the hadronic recoil and the last one is for the uncertainty due to the electron energy response.}
    		\label{tab:syst_var_Matrix}
    		\resizebox{0.55\textwidth}{!}{\tiny
    			\begin{tabular}{lcccccc}
    				\hline
    				\multicolumn{7}{c}{Systematic Variation No.~1} \\
    				$p_T^W$ bin	& 0--2 GeV		& 2--5 GeV		& 5--8 GeV		& 8--11 GeV	& 11--15 GeV	& 15--600 GeV	\\
    				$0<u_T<2$ & 0.1876 & 0.1738 & 0.1196 & 0.0715 & 0.0363 & 0.0071 \\ 
    				$2<u_T<5$ & 0.4642 & 0.4588 & 0.4109 & 0.3120 & 0.2022 & 0.0456 \\ 
    				$5<u_T<8$ & 0.2382 & 0.2503 & 0.2938 & 0.3388 & 0.3107 & 0.1112 \\ 
    				$8<u_T<11$ & 0.0777 & 0.0840 & 0.1227 & 0.1822 & 0.2535 & 0.1644 \\ 
    				$11<u_T<15$ & 0.0272 & 0.0275 & 0.0439 & 0.0780 & 0.1503 & 0.2216 \\
    				\multicolumn{7}{c}{Systematic Variation No.~2} \\
    				$p_T^W$ bin	& 0--2 GeV		& 2--5 GeV		& 5--8 GeV		& 8--11 GeV	& 11--15 GeV	& 15--600 GeV	\\
    				$0<u_T<2$ & 0.1754 & 0.1669 & 0.1193 & 0.0720 & 0.0356 & 0.0070 \\ 
    				$2<u_T<5$ & 0.4665 & 0.4607 & 0.4091 & 0.3144 & 0.2009 & 0.0457 \\ 
    				$5<u_T<8$ & 0.2410 & 0.2506 & 0.2957 & 0.3323 & 0.3113 & 0.1137 \\
    				$8<u_T<11$ & 0.0834 & 0.0880 & 0.1231 & 0.1838 & 0.2511 & 0.1667 \\
    				$11<u_T<15$ & 0.0280 & 0.0281 & 0.0437 & 0.0788 & 0.1532 & 0.2209 \\
    				\multicolumn{7}{c}{Systematic Variation No.~3} \\
    				$p_T^W$ bin	& 0--2 GeV		& 2--5 GeV		& 5--8 GeV		& 8--11 GeV	& 11--15 GeV	& 15--600 GeV	\\
    				$0<u_T<2$ & 0.1776 & 0.1702 & 0.1200 & 0.0698 & 0.0340 & 0.0067 \\ 
    				$2<u_T<5$ & 0.4647 & 0.4618 & 0.4098 & 0.3203 & 0.1988 & 0.0442 \\ 
    				$5<u_T<8$ & 0.2393 & 0.2496 & 0.2967 & 0.3359 & 0.3078 & 0.1121 \\ 
    				$8<u_T<11$ & 0.0850 & 0.0852 & 0.1222 & 0.1802 & 0.2584 & 0.1630 \\ 
    				$11<u_T<15$ & 0.0273 & 0.0275 & 0.0428 & 0.0762 & 0.1542 & 0.2245 \\ 
    				\multicolumn{7}{c}{Systematic Variation No.~4} \\
    				$p_T^W$ bin	& 0--2 GeV		& 2--5 GeV		& 5--8 GeV		& 8--11 GeV	& 11--15 GeV	& 15--600 GeV	\\
    				$0<u_T<2$ & 0.1815 & 0.1744 & 0.1215 & 0.0730 & 0.0366 & 0.0068 \\ 
    				$2<u_T<5$ & 0.4612 & 0.4577 & 0.4110 & 0.3157 & 0.2022 & 0.0467 \\ 
    				$5<u_T<8$ & 0.2440 & 0.2505 & 0.2941 & 0.3311 & 0.3114 & 0.1126 \\ 
    				$8<u_T<11$ & 0.0811 & 0.0842 & 0.1209 & 0.1817 & 0.2509 & 0.1641 \\ 
    				$11<u_T<15$ & 0.0263 & 0.0279 & 0.0438 & 0.0799 & 0.1504 & 0.2199 \\ 
    				\multicolumn{7}{c}{Systematic Variation No.~5} \\
    				$p_T^W$ bin	& 0--2 GeV		& 2--5 GeV		& 5--8 GeV		& 8--11 GeV	& 11--15 GeV	& 15--600 GeV	\\
    				$0<u_T<2$ & 0.1808 & 0.1697 & 0.1199 & 0.0707 & 0.0355 & 0.0067 \\ 
    				$2<u_T<5$ & 0.4623 & 0.4617 & 0.4129 & 0.3213 & 0.1973 & 0.0443 \\ 
    				$5<u_T<8$ & 0.2424 & 0.2498 & 0.2940 & 0.3354 & 0.3130 & 0.1121 \\ 
    				$8<u_T<11$ & 0.0818 & 0.0857 & 0.1212 & 0.1792 & 0.2526 & 0.1676 \\ 
    				$11<u_T<15$ & 0.0274 & 0.0277 & 0.0422 & 0.0760 & 0.1561 & 0.2229 \\ 
    				\multicolumn{7}{c}{Systematic Variation No.~6} \\
    				$p_T^W$ bin	& 0--2 GeV		& 2--5 GeV		& 5--8 GeV		& 8--11 GeV	& 11--15 GeV	& 15--600 GeV	\\
    				$0<u_T<2$ & 0.1740 & 0.1716 & 0.1241 & 0.0739 & 0.0364 & 0.0066 \\
    				$2<u_T<5$ & 0.4625 & 0.4609 & 0.4116 & 0.3207 & 0.2011 & 0.0462 \\ 
    				$5<u_T<8$ & 0.2446 & 0.2489 & 0.2917 & 0.3303 & 0.3145 & 0.1113 \\ 
    				$8<u_T<11$ & 0.0857 & 0.08433 & 0.1210 & 0.1817 & 0.246 & 0.1649 \\ 
    				$11<u_T<15$ & 0.0280 & 0.0287 & 0.0429 & 0.0758 & 0.1537 & 0.2216 \\
    				\multicolumn{7}{c}{Systematic Variation No.~7} \\
    				$p_T^W$ bin	& 0--2 GeV		& 2--5 GeV		& 5--8 GeV		& 8--11 GeV	& 11--15 GeV	& 15--600 GeV	\\
    				$0<u_T<2$ & 0.1803 & 0.1725 & 0.1233 & 0.0711 & 0.0352 & 0.0071 \\ 
    				$2<u_T<5$ & 0.4648 & 0.4612 & 0.4121 & 0.3197 & 0.2025 & 0.0454 \\ 
    				$5<u_T<8$ & 0.2423 & 0.2507 & 0.2934 & 0.3320 & 0.3110 & 0.1092 \\ 
    				$8<u_T<11$ & 0.0810 & 0.0832 & 0.1188 & 0.1826 & 0.2545 & 0.1643 \\ 
    				$11<u_T<15$ & 0.0263 & 0.0268 & 0.0434 & 0.0768 & 0.1493 & 0.2239 \\ 
    				\multicolumn{7}{c}{Systematic Variation No.~8} \\
    				$p_T^W$ bin	& 0--2 GeV		& 2--5 GeV		& 5--8 GeV		& 8--11 GeV	& 11--15 GeV	& 15--600 GeV	\\
    				$0<u_T<2$ & 0.1805 & 0.1722 & 0.1218 & 0.0705 & 0.0379 & 0.0070 \\ 
    				$2<u_T<5$ & 0.4648 & 0.4602 & 0.4123 & 0.3172 & 0.2052 & 0.0466 \\ 
    				$5<u_T<8$ & 0.2399 & 0.2481 & 0.2927 & 0.3379 & 0.3114 & 0.1137 \\ 
    				$8<u_T<11$ & 0.0826 & 0.0863 & 0.1215 & 0.1805 & 0.2477 & 0.1653 \\ 
    				$11<u_T<15$ & 0.0266 & 0.0278 & 0.0432 & 0.0764 & 0.1517 & 0.2235 \\ 
    				\multicolumn{7}{c}{Systematic Variation No.~9} \\
    				$p_T^W$ bin	& 0--2 GeV		& 2--5 GeV		& 5--8 GeV		& 8--11 GeV	& 11--15 GeV	& 15--600 GeV	\\
    				$0<u_T<2$ & 0.1774 & 0.1709 & 0.1241 & 0.0717 & 0.0348 & 0.0064 \\ 
    				$2<u_T<5$ & 0.4618 & 0.4563 & 0.4077 & 0.3188 & 0.1980 & 0.0445 \\ 
    				$5<u_T<8$ & 0.2444 & 0.2525 & 0.2958 & 0.3335 & 0.3138 & 0.1116 \\ 
    				$8<u_T<11$ & 0.0833 & 0.0866 & 0.1216 & 0.1798 & 0.2512 & 0.1657 \\ 
    				$11<u_T<15$ & 0.0275 & 0.0278 & 0.0417 & 0.0782 & 0.1542 & 0.2226 \\
    				\multicolumn{7}{c}{Systematic Variation No.~10} \\
    				$p_T^W$ bin	& 0--2 GeV		& 2--5 GeV		& 5--8 GeV		& 8--11 GeV	& 11--15 GeV	& 15--600 GeV	\\
    				$0<u_T<2$ & 0.1826 & 0.1720 & 0.1198 & 0.0708 & 0.0370 & 0.0073 \\ 
    				$2<u_T<5$ & 0.4598 & 0.4584 & 0.4100 & 0.3168 & 0.2026 & 0.0469 \\ 
    				$5<u_T<8$ & 0.2420 & 0.2483 & 0.2988 & 0.3346 & 0.3091 & 0.1120 \\ 
    				$8<u_T<11$ & 0.0827 & 0.0876 & 0.1195 & 0.1819 & 0.2494 & 0.1628 \\ 
    				$11<u_T<15$ & 0.0273 & 0.0278 & 0.0430 & 0.0774 & 0.1546 & 0.2204 \\ 
    				\multicolumn{7}{c}{Systematic Variation No.~11} \\
    				$p_T^W$ bin	& 0--2 GeV		& 2--5 GeV		& 5--8 GeV		& 8--11 GeV	& 11--15 GeV	& 15--600 GeV	\\
    				$0<u_T<2$ & 0.1790 & 0.1707 & 0.1192 & 0.0716 & 0.0349 & 0.0072 \\ 
    				$2<u_T<5$ & 0.4624 & 0.4629 & 0.4102 & 0.3176 & 0.2030 & 0.0472 \\
    				$5<u_T<8$ & 0.2436 & 0.2484 & 0.2967 & 0.3341 & 0.3116 & 0.1108 \\ 
    				$8<u_T<11$ & 0.0839 & 0.0853 & 0.1223 & 0.1830 & 0.2483 & 0.1653 \\
    				$11<u_T<15$ & 0.0259 & 0.0271 & 0.0431 & 0.0763 & 0.1561 & 0.2229 \\
    				\hline
    		\end{tabular}}
    	\end{center}
    \end{table*}

   \begin{table*}[!]
	\begin{center}
		\caption{\small The fiducial correction $F(u_T)$ in each $u_T$ bin for the eleven systematic variations. The fiducial correction is the probability to pass the particle level selection for the events that pass theion selection. The first ten systematic variations are for the uncertainty due to the hadronic recoil and the last one is for the uncertainty due to the electron energy response.}
		\label{tab:syst_var_Fiducial}
		\begin{tabular}{lccccc}
			\hline
			$u_T$ bin	& 0--2 GeV		& 2--5 GeV		& 5--8 GeV		& 8--11 GeV	& 11--15 GeV		\\
			\hline
			Systematic Variation No.~1 &0.8639 & 0.8705 & 0.8778 & 0.8814 & 0.9011 \\
			Systematic Variation No.~2 &0.8629 & 0.8686 & 0.8787 & 0.8817 & 0.9033 \\ 
			Systematic Variation No.~3 &0.8612 & 0.8703 & 0.8796 & 0.8824 & 0.9003 \\ 
			Systematic Variation No.~4 &0.8637 & 0.8673 & 0.8789 & 0.8819 & 0.9002 \\ 
			Systematic Variation No.~5 &0.8637 & 0.8690 & 0.8803 & 0.8795 & 0.9037  \\ 
			Systematic Variation No.~6 &0.8638 & 0.8686 & 0.8779 & 0.8799 & 0.9020 \\ 
			Systematic Variation No.~7 &0.8634 & 0.8691 & 0.8805 & 0.8830 & 0.8996 \\ 
			Systematic Variation No.~8 &0.8651 & 0.8695 & 0.8795 & 0.8821 & 0.8992 \\ 
			Systematic Variation No.~9 &0.8664 & 0.8691 & 0.8800 & 0.8819 & 0.9004 \\ 
			Systematic Variation No.~10 &0.8630 & 0.8691 & 0.8786 & 0.8808 & 0.9007 \\ 
			Systematic Variation No.~11 &0.8615 & 0.8700 & 0.8798 & 0.8842 & 0.9004 \\ 
			\hline
		\end{tabular}
		
	\end{center}
\end{table*}

    As a validation, the $\chi^2$ values calculated from the fast folding approach are compared to those provided in Table \ref{tab:result_chi2}. The background-subtracted data is fluctuated with the statistical uncertainty from the data in order to estimate the impact on $\chi^2/$ndf from the data statistics. The difference between the chi-square values calculated from the PMCS simulation and that calculated from the fast folding is negligible compared to the impact of the statistical fluctuation of the data, hence validating this approach. 
    


\end{document}